# *Atomistic Description of Spin Crossover Under Pressure and its Giant Barocaloric Effect*

Sergi Vela,[a,b,*] Jordi Ribas-Arino,[b] Steven P. Vallone[c,d], António M. dos Santos[e], Malcolm A. Halcrow[f] and Karl Sandeman[c,d]

[a]*Institut de Química Avançada de Catalunya (IQAC-CSIC), Barcelona, Spain.* sergi.vela@iqac.csic.es

[b]*Departament de Ciència de Materials i Química Física and IQTCUB, Universitat de Barcelona, Martí i Franquès 1, E-08028, Barcelona, Spain.*

[c]*Department of Physics, Brooklyn College of The City University of New York, 2900 Bedford Avenue Brooklyn NY 11210, USA*
[d]*Physics Program, The Graduate Center, CUNY, 365 Fifth Avenue NY 10016, USA*

[e] *Neutron Sciences Directorate, Oak Ridge National Laboratory, Oak Ridge, TN 37831-6393, USA*

[f] *School of Chemistry, University of Leeds, Woodhouse Lane, Leeds LS2 9JT, UK*

**Abstract**

The pressure-dependent evolution of the Spin Crossover (SCO) transition has garnered significant interest due to its connection to the giant barocaloric effect (BCE) near room temperature. Pressure alters both the molecular and solid-state structures of SCO materials, affecting the relative stability of low- and high-spin states and, consequently, the transition temperature ($T_{1/2}$). Crucially, the shape of the $T_{1/2}$ vs. pressure curve dictates the magnitude of the BCE, making its accurate characterization essential for identifying high-performance materials. In this work, we investigate the nonlinear $T_{1/2}$ vs. pressure behavior of the prototypical SCO complex [FeL$_2$][BF$_4$]$_2$ [L = 2,6-di(pyrazol-1-yl)pyridine] using solid-state PBE+U computations. Our results unveil the mechanisms by which pressure influences its SCO transition, including the onset of a phase transition, as well as the key role of low-frequency phonons in the BCE. Furthermore, we establish a computational protocol for accurately modeling the BCE in SCO crystals, providing a powerful tool for the rapid and efficient discovery of new materials with enhanced barocaloric performance.

**1. Introduction**

Spin crossover (SCO) complexes undergo a controlled phase transition between a high-spin (HS) state and a low-spin (LS) state with different physical properties.[1–8] The switch can be triggered by several stimuli, being temperature and light the most common.[9–11] This capacity to switch their behavior, coupled with the different physical properties of the HS and LS states, made SCOs candidate components in displays, memories, sensors, and actuators.[12–16] Recently, the presence of a giant barocaloric effect (BCE) near room temperature has been predicted,[17,18] and then demonstrated for a number of SCO compounds,[19–22] while more recently a giant elastocaloric effect has been observed in a SCO/polymer blend.[23] The barocaloric and elastocaloric potential of SCO compounds stems from the large HS-LS entropy difference at the phase transition.[24,25] When temperature is used to trigger the switch, and the sample is heated to the so-called transition temperature ($T_{1/2}$), the SCO releases heat. If a stimulus is applied adiabatically (instead of isothermally), there is a change in temperature labelled $\Delta T_{ad}$. This metric is one measure of the cooling capacity of the material associated with the SCO. In the case of the barocaloric effect, $\Delta T_{ad}$ depends on the applied pressure, and its maximum is connected with how much $T_{1/2}$ changes with pressure ($dT_{1/2}/dP$), at a given applied pressure.[17,26,27] Single $dT_{1/2}/dP$ values have been reported in the scarce SCO literature on the topic, resulting from a near-constant increase of $T_{1/2}$ with $P$.[20] One exception to the rule is the SCO complex [FeL$_2$][BF$_4$]$_2$ [L = 2,6-di(pyrazol-1-yl)pyridine] (**1**).[19,28]

For this compound, magnetization measurements revealed that the evolution of its $T_{1/2}$ with pressure is non-linear.[19] Three regimes were identified: (i) a low-pressure regime at which $T_{1/2}$ increases slowly with pressure ($dT_{1/2}/dP \approx 100$ K·GPa$^{-1}$), (ii) a mid-pressure regime (above 200 MPa), in which $T_{1/2}$ becomes more susceptible to pressure

($dT_{1/2}/dP \approx 400$ K·GPa$^{-1}$), and (iii) a high-pressure regime (above 280 MPa) at which point we recover the stiffer response to pressure ($dT_{1/2}/dP \approx 200$ K·GPa$^{-1}$). These can be labelled regions I, II and III, respectively. In the low-pressure regime, an irreversible $\Delta T_{ad}$ value of 2 K (4.5 K) was achieved at 19 MPa (43 MPa).[19] Shortly after, another SCO complex with strong BCE was reported, with a reversible $\Delta T_{ad}$ value of 10 K at 55 MPa.[20] It is thus clear that the large library of reported SCO complexes might be of great interest for the development of new materials for BCE. Prior to any discovery endeavor, however, it is convenient to understand how pressure impacts the spin state energetics of SCO complexes and, for this task, atomistic computational tools can offer unparalleled insight.[21]

In this manuscript, we provide some insight into this complex topic by studying the pressure-dependence of the spin crossover transition in **1**. Compound **1** has been recently studied theoretically using a mean-field model Hamiltonian.[29] The model used data from DFT calculations of the isolated SCO molecule, and was fitted to the experimental calorimetry data of the SCO transition at 43 MPa. The study revealed that low-frequency (i.e. optical) phonons contribute to most of the entropy change associated with the BCE. To expand this study, here we use atomistic methods to investigate the evolution of the solid-state structure and lattice phonons of **1** with pressure, as well as the thermodynamic contributions affecting its spin-state energy balance. In so doing, we identify the reasons why crystals of **1** present different regimes of $dT_{1/2}/dP$, and why SCO compounds in general tend to present an increase of $T_{1/2}$ with pressure. Our results thus advance the understanding of pressure-induced spin crossover behavior, and help establishing a computational framework to tackle the modeling of the BCE in SCO crystals, and the enhanced exploration of potentially better candidates.

## 2. Methodology and Computational Details

***Energetics of the SCO transition.*** The SCO transition is studied here using the Gibbs free energy difference at a given external pressure ($P$) and temperature ($T$) (see eq. 1).

$$\Delta G(T,P) = \Delta H_{elec}(P) + \Delta H_{vib}(T,P) \\ - T(\Delta S_{vib}(T,P) + \Delta S_{elec}) \\ + P\Delta V(P) \quad (1)$$

with: $\Delta X = X_{HS} - X_{LS}$

where $\Delta H_{elec}(P)$ is the adiabatic energy difference between the HS and LS phases, and $\Delta S_{elec}$ accounts for the difference in the number of electronic configurations between both phases ($2S + 1$). Considering quintuplet ($S = 2$) and singlet ($S = 0$) multiplicities, and weak intermolecular coupling, $\Delta S_{elec}$ is 13.38 J·K$^{-1}$·mol$^{-1}$ at all pressures. The vibrational terms $\Delta H_{vib}(T,P)$ and $\Delta S_{vib}(T,P)$ account for the changes in the vibrational energy levels of the molecule, as well as in lattice phonons. For simplicity, both are labelled *phonons* in this manuscript. Vibrational terms are modelled using the corresponding harmonic oscillator expressions (see equations 2-3), together with the frequency of the computed phonons ($v_i$).

$$H_{vib} = \sum_{i}^{N_{vib}} \left( \frac{1}{2}hv_i + \frac{hv_i e^{-hv_i/k_B T}}{1 - e^{-hv_i/k_B T}} \right) \quad (2)$$

$$S_{vib} = \sum_{i}^{N_{vib}} \left( \frac{hv_i}{T} \frac{1}{e^{hv_i/k_B T} - 1} - k_B \ln\left(1 - e^{-hv_i/k_B T}\right) \right) \quad (3)$$

Finally, the term $P\Delta V(P)$ accounts for the impact of pressure on volume changes at the SCO transition. At ambient (or low) pressure, this term is negligible, but at the pressures considered here, its contribution to the spin transition cannot be ignored.

Considering the terms in equation 1, at the spin transition ($\Delta G = 0$), the transition temperature ($T_{1/2}$) is:

$$T_{1/2}(P) = \frac{\Delta H_{elec}(P) + \Delta H_{vib}(T_{1/2}, P) + P\Delta V(P)}{\Delta S_{vib}(T_{1/2}, P) + \Delta S_{elec}} \quad (4)$$

**Energy decomposition:** For analysis purposes, $H_{elec}$ has been decomposed as a sum of one and two-body terms, describing the energies of the isolated species (SCO and counterions, CI), as well as their intermolecular interaction energies (eq. 5).

$$H_{elec}^i = H_{SCO}^i + H_{CI}^i + H_{SCO-SCO}^i + H_{SCO-CI}^i \\ + H_{CI-CI}^i \quad (5)$$

with $i$ = HS/LS. For instance, $H_{SCO}^{HS}$ is the sum of SCO molecular energies at the HS phase, and $\Delta H_{SCO}$ is their difference between the HS and LS phases. All terms were obtained from single-point calculations of the optimized LS and HS unit cells, as discussed elsewhere.[30,31]

**Computational details:** All computations were performed with Quantum Espresso[32] using the PBE exchange–correlation functional,[33] spin unrestricted formalism, D3-BJ dispersion corrections,[34–37] and Vanderbilt ultrasoft pseudopotentials.[38] A 2x2x1 Monkhorst-Pack mesh was used to sample the irreducible part of the Brillouin zone. The Hubbard-like $U$ parameter[39,40] for Fe$^{II}$ was set to 2.35 eV, according to a previous benchmark study.[41]

The minimum energy structures of the HS and LS phases of **1**, for each different value of external compressive pressure (0-500 MPa), were obtained by performing successive variable-cell geometry relaxations, in which the lattice parameters as well as the atomic positions were optimized simultaneously until the atomic forces were smaller than $2.0·10^{-4}$ a.u., and the difference between the imposed and actual cell pressures was smaller than 0.01 kbar (1 MPa). In these calculations, the number of plane waves has been kept constant at a kinetic energy cutoff of 70 Ry for the wavefunction (ecutwfc) and of 700 Ry for the charge density (ecutrho). The spin state of the iron atoms is defined in the initial configuration, and maintained along the optimization. The initial HS and LS unit cells at 20 MPa were taken from ambient-pressure experimental crystal structures.[28] For subsequent pressures, the optimized structures from the previous run were used as starting points. All unit cells contain two SCO molecules and four counterions. All energy values are given *per* mol of **1**. After the variable-cell relaxations, the resulting structures went through subsequent fixed-cell optimizations, in which ecutwfc was reduced to 35 Ry and the force convergence threshold was tightened to $1.0·10^{-5}$ a.u. The final fixed-cell optimized structures were then used to calculate the frequencies ($\nu_i$) of the normal modes for both phases at each value of external pressure. The frequencies were calculated using a finite-difference approach, with a step size of 0.01 atomic units, and Γ-point sampling of the first Brillouin zone, and were then employed to calculate the relevant thermodynamic quantities by means of Equations 2 and 3.

### 3. Results and Discussion

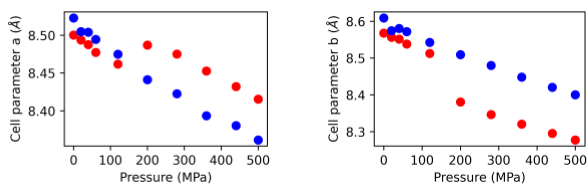

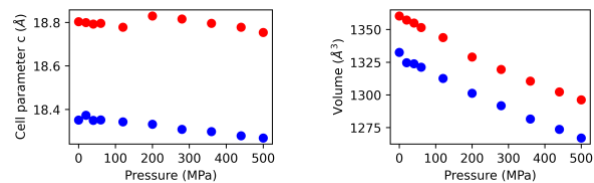

**Figure 1.** Evolution of volume and unit cell parameters with pressure for the HT (red) and LT (blue) phases of **1**.

**Structural features.** The optimized unit cells of **1** show the expected compression with pressure, with a *ca.* 4.91% and 4.72% of volume compression at 500 MPa (with respect to ambient pressure) for the LT and HT phases of **1**, respectively. Interestingly, while the compression of the LT phase is monotonic, that of the HT phase shows a discontinuity between 120 and 200 MPa, which manifests itself in two ways: first, as a sharp increase along *a* and *c* unit cell parameters, accompanied by a decrease along b, resulting in subtle changes in the overall volume, see Figure 1). Second, in internal changes in the geometry of the SCO and counterion molecules (see Figure S2). These results suggest that a phase transition occurs in the HT phase of **1** between 120 and 200 MPa. It is not possible to compare these results with the existing neutron diffraction data due to temperature effects being absent in our optimized unit cells, and also due to the sparsity of experimental data.[19]

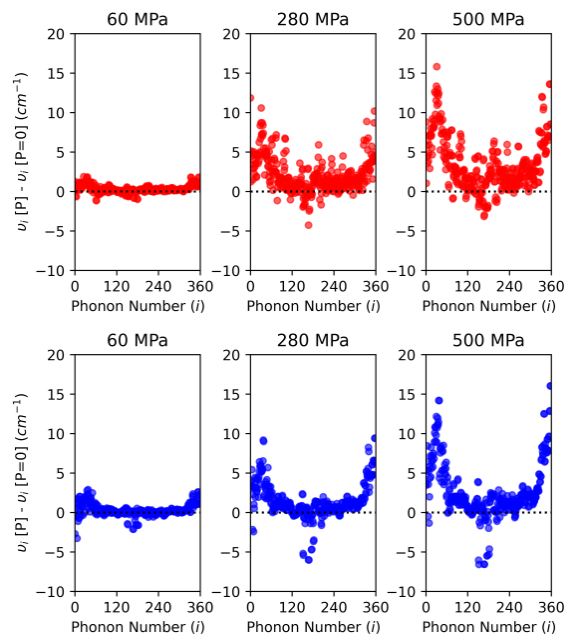

**Figure 2.** Evolution of HS (top, red) and LS (bottom, blue) phonon frequencies when increasing pressure. Each dot represents the difference between the frequency of a phonon at the selected pressures (60, 280 and 500 MPa) and at ambient pressure. For more pressure values, see Figure S3. For an analysis of the associated eigenvectors, see Figure 4.

**Phonons.** The application of external pressure modifies the frequencies of phonons in crystals of **1**. We consistently observe a U-shaped frequency increase as pressure ramps up (see Figures 2 and S3). Basically, the frequency of extreme low- and high-frequency modes gradually increases with pressure, in a similar fashion for both spin states. Such an increase in frequency is somehow expected; molecules should be increasingly confined at higher pressures. However, it is surprising that such an increase is not consistent for all phonons and, indeed, the frequency of some of them even decreases when pressure is applied (see Figures 2 and S3). It is thus clear that applied pressure does not impact the phonons uniformly, nor focuses on a specific range of phonons.

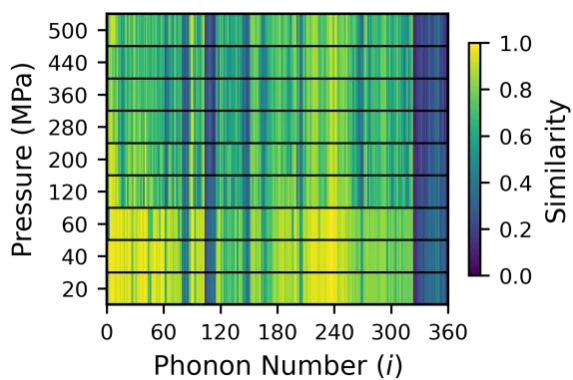

**Figure 3.** Similarity of HS phonon eigenvectors with respect to ambient pressure. Large values indicate that the eigenvector of the $i$-th phonon at a given pressure is very similar to the same $i$-th phonon at ambient pressure. Small values indicate either a change in character, or a change in phonon ordering due to a frequency shift. See Figure S4 for LS and for more details.

Despite the change in phonon frequencies with pressure, their *character* (i.e. its eigenvector) does not significantly change. It defines which molecules are involved, and which is their motion, so it is worth analyzing if changes occur under an external pressure. Phonon character changes have been quantified using a similarity metric described in section S4 of the ESI, and taking the ambient-pressure phonons as the reference. Similarities close to 1.0 are retrieved for most phonons at low external pressures (20-60 MPa, see Figure 3), meaning that pressure barely changes their character. The main exception are the high-frequency HS (see Figure 3) and LS phonons (see Figure S4). This is likely due to a change in phonon ordering as a result of the frequency shift we observed in Figure 2, coupled with the very localized nature of these modes. At higher pressures at and above 120 MPa, the impact of the uncovered phase transition manifests itself once again, with similarities gradually dropping to 0.6-0.8, indicating a slight change in phonon character. Considering that the Hessians have been computed independently -and numerically- at each pressure, this analysis is helpful to ensure that meaningful results have been obtained, as one can confidently assign each phonon to its main chemical motif and character. In our opinion, these features indicate that our implementation and use of the finite differences' technique for the phonon computations with Quantum Espresso is satisfactory.

**Spin-state energies.** The pressure evolution of all terms contributing to the spin state energy balance between the HT and LT phases of **1** (see eq. 4) is collected in Figure 4. The adiabatic energy difference ($\Delta H_{elec}$) remains quite constant during the whole range of pressures, ranging from 20.6 kJ/mol at ambient pressure, to 19.5 kJ/mol at 500 MPa. Such a decrease in the relative stability of the LT phase is compensated by the $P\Delta V$ term, which penalizes the HT phase due to its larger unit cell volume, and reaches 4.4 kJ/mol at 500 MPa. We must thus highlight the importance of the $P\Delta V$ term to describe the SCO transition at high pressures, which justifies its addition (with respect to previous work[42,43]) to equation **1**.

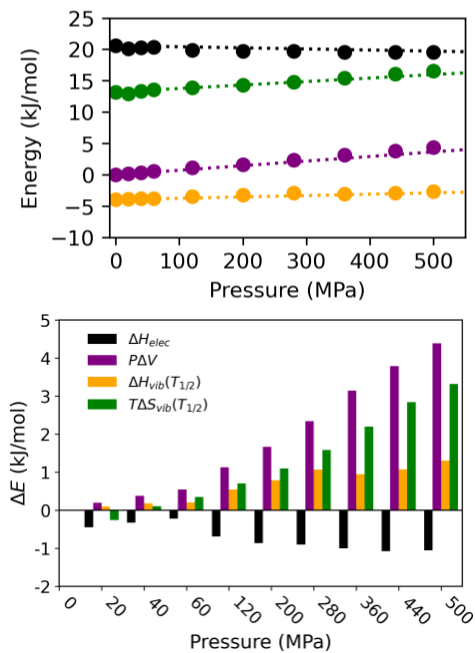

**Figure 4.** (top) Evolution of $\Delta H_{elec}$ (**black**), $P\Delta V$ (purple), $\Delta H_{vib}(T_{1/2})$ (orange) and $T_{1/2} \cdot \Delta S_{vib}(T_{1/2})$ (green). (bottom) Relative difference of each term with respect to ambient pressure. Positive (negative) values indicate an overestabilization of the LS (HS) state, and thus an increase (decrease) of $T_{1/2}$.

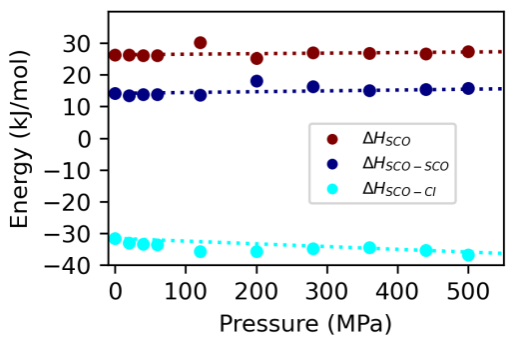

**Figure 5.** Evolution of the main energy terms decomposing $\Delta H_{elec}$ as defined in eq 5. Terms $\Delta H_{CI}$ and $\Delta H_{CI-CI}$ are close to 0.0 at all range of pressures and are not shown. Dashed lines are drawn between the extreme values of each serie to better visualize trends.

To further investigate the unit cell contributions to $\Delta H_{elec}$, we used a decomposition scheme reported in the computational details, in which molecular contributions are evaluated with one- and two-body terms, and grouped by molecular type (SCO and counterion in this case). This decomposition scheme has been used in previous work to decipher crystal packing effects in SCO crystals.[30,31] In our work, the energies of the isolated cations ('SCO') and anions ('BF$_4$'), as well as their intermolecular interaction energies, were evaluated for all optimized HS and LS unit cells of **1** (see Figure 5). The evolution of these terms with pressure enables us to understand their impact on the stability of specific molecules in the unit cell. At 120 MPa, the increase in $\Delta H_{SCO}$ is due to an increase of $H_{SCO}^{HS}$, which indicates that the SCO molecules in the HS state are destabilized. This is likely related with the slight change in the evolution of $\bar{d}$(Fe-N) and $\Sigma$ for the HS state at this pressure (see Figure S2) ($\Sigma$ is a measure of the distortion of the coordination sphere from a perfect octahedron[5,44]). Overall, this instability likely triggers a phase transition in the HS phase observed between 120 and 200 MPa in the cell parameters (see Figure **1**). After completing this phase transition (for $P \geq 200$ MPa), the entire energy balance changes; the increase in $\Delta H_{SCO}$ is compensated by smaller $\Delta H_{SCO-SCO}$ and $\Delta H_{SCO-CI}$ values, which suggests a slight rearrangement of the molecules in the crystal, but the overall $\Delta H_{elec}$ evolves smoothly with pressure. Besides those changes, the remaining terms (related to the counterions) are rather unaltered by pressure, as expected.

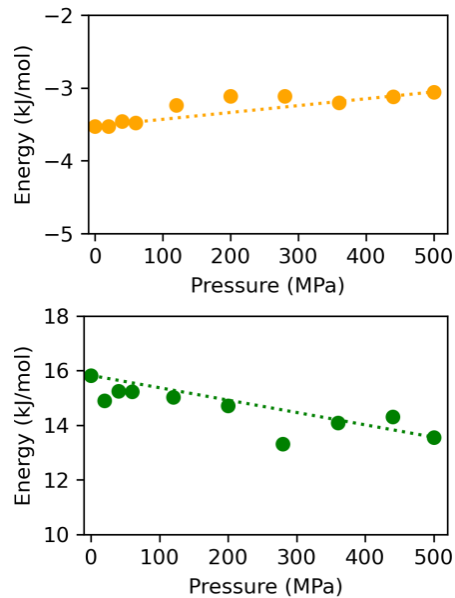

**Figure 6.** Evolution of $\Delta H_{vib}(T)$ (orange) and $T\Delta S_{vib}(T)$ (green) with pressure, using $T$=300K. Dashed lines are drawn between extreme values to better visualize trends

The contribution of the vibrational terms $\Delta H_{vib}$ and $\Delta S_{vib}$ to the spin crossover transition is more complex. In general, at moderate temperatures all phonons contribute to $\Delta H_{vib}$, while only low-frequency ones contribute to $\Delta S_{vib}$.[45] Thus, given that pressure affects all phonons (see Figure S2 and discussion above), we can expect changes to both terms when applying pressure. Quantitatively, $\Delta H_{vib}$ always favors the HS state (negative value, see Figure 4-top), but its absolute value decreases with pressure (positive delta, see Figure 4-bottom). In other words, pressure diminishes the HS-stabilizing effect of $\Delta H_{vib}$. In turn, $\Delta S_{vib}$ follows the opposite behavior. $\Delta S_{vib}$ always favors the LS state, and its absolute value diminishes with pressure, even if it does not seem so in Figure 4.

The reason why Figure 4 shows an increase of $T_{1/2}\Delta S_{vib}(T_{1/2})$ with pressure is because $T_{1/2}$ is increasing as a result of the changes in all terms combined. Rather, if we monitor the $\Delta S_{vib}$ term at a constant temperature, for instance at 300 K, we can clearly see a regular decrease, except at 280 MPa, where we observe an outlier that deviates several kJ/mol (see Figure 6). The same analysis for $\Delta H_{vib}$ also reveals a different trend at moderate (120-280 MPa) than at higher (above 280 MPa) pressures, although with a much smaller energetic impact. This change is likely associated with the onset of the phase transition that we identified above between 120 and 200 MPa. Overall, these trends are the result of the changes in the phonon frequencies of the HS and LS states. Unfortunately, we could not identify which group of individual phonons is responsible for this pattern, and thus, which molecular features are able to explain the outliers in the behavior of $\Delta H_{vib}$ and $\Delta S_{vib}$ with pressure. However, considering the much larger impact on $\Delta S_{vib}$ than on $\Delta H_{vib}$, we may hypothesize that the responsibility for the existence of three regimes relies on a group of low-frequency lattice phonons. This, together with previous related work,[29] highlights the importance of low-frequency phonons to the BCE.

**Transition Temperature.** Using equation 4, we constructed our computational prediction for the $T_{1/2}$ vs. $P$ curve of **1** (see Figure 7). Not only the overall trend is captured, but only a small deviation of 5-20 K appears between computed and experimental $T_{1/2}$ values at most pressures. Such a deviation is indeed excellent, and stems from our thorough benchmark of the applied methodology in previous work for SCO systems, which often included compound **1**.[41,42,46] We must note two relevant sources of disagreement. First, the experimental $T_{1/2}$ measurements in the low-pressure regime could be affected by the thermal hysteresis of **1**. Second, the strange evolution of $T_{1/2}$ at -or above- 280 MPa, which we ascribed to the $\Delta H_{vib}$ and -specially- $\Delta S_{vib}$ trends discussed above.

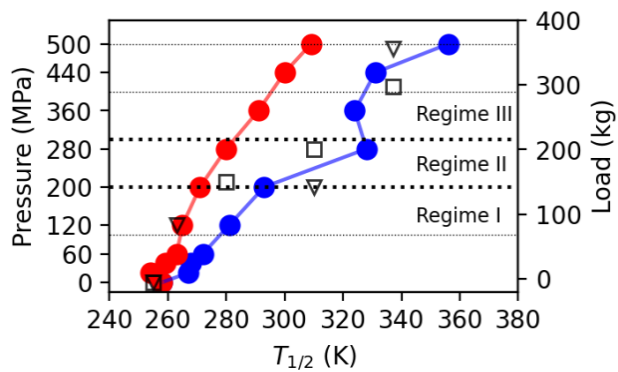

**Figure 7.** Comparison between experimental[19] (**black**) and computed spin crossover transition temperature ($T_{1/2}$) using equation 4 (blue). In red, the computed $T_{1/2}$ using ambient pressure $\Delta H_{vib}$ and $\Delta S_{vib}$ terms at all pressures. That is, removing the impact of phonon changes due to pressure (see text). Two experimental data is represented in terms of applied pressure (triangles, left axis) or applied load (squares, right axis).

From the computed $T_{1/2}$ vs. $P$ curves we obtain $dT_{1/2}/dP$ figures of 175, 437 and 127 K·GPa$^{-1}$ for regimes I, II and III, respectively. The anomaly at or above 280 MPa means that $dT_{1/2}/dP$ is likely overestimated for regime II and underestimated for regime III. More importantly, our results indicate that the existence of three different regimes originates mainly in the non-linear behavior of the vibrational terms ($\Delta H_{vib}$ and $\Delta S_{vib}$) responsible for the spin transition. This becomes evident when comparing the predicted $T_{1/2}$ vs. $P$ curve with, and without, accounting the changes in vibrational contributions (see Figure 7). When using ambient pressure vibrational terms ($\Delta H_{vib}(T_{1/2}, P = 0)$ and $\Delta S_{vib}(T_{1/2}, P = 0)$) at all pressures, we observe a constant increase of $T_{1/2}$ (similar to other SCO systems under pressure[20]), which simply follows the combined effect of the $P\Delta V$ and $\Delta H_{elec}$ terms (see Figure 4 and discussion above). Finally, we ascribe the onset of an outlier $T_{1/2}$ value at 280 MPa to an unexpectedly low $\Delta S_{vib}$ term at this pressure, which must originate in pressure-driven changes in the low-frequency (i.e. lattice) phonons of **1**.

## 4. Conclusions

Solid-state DFT computations have been carried out to study the impact of an external pressure on the spin transition of **1**. The non-linear behavior of $T_{1/2}$ with pressure is captured, and the three regimes of $dT_{1/2}/dP$ found experimentally are naturally reproduced. The accuracy of the computed $T_{1/2}$ curve is remarkable, which enables us to extract performance metrics relevant to the BCE. The irregular evolution of $dT_{1/2}/dP$ with pressure (*i.e.* the three regimes) stems mainly from the non-linear evolution of vibrational terms ($\Delta H_{vib}$ and $\Delta S_{vib}$), which itself originates in the non-linear evolution the phonon frequencies when pressure is applied. Indeed, our solid-state computations indicate a frequency increase for both low- (optical) and high-frequency (acoustic) phonons with pressure, in contrast with DFT data from isolated molecules.[29] We also found evidence of a phase transition occurring on the HS crystals of **1** between 120 and 200 MPa, which contributes to this non-linear behavior. On the methodological side, we stress the efficiency and accuracy of the computational framework, and we highlight the importance of the $P\Delta V$ term to describe the energetics of the spin transition at this pressure conditions. We hope that this study may help researchers establish a strategy aimed at the systematic discovery of new SCO complexes of potential interest in barocalorics using atomistic methods.

## Data Availability

Data for this article, including the optimized unit-cell structures for the HS and LS states of **1** under external pressures, as well as the associated geometry optimization computations, are available at 10.5281/zenodo.15641001. Temporary access link for reviewers:

https://zenodo.org/records/15641001?preview=1&token=eyJhbGciOiJIUzUxMiJ9.eyJpZCI6IjMxNDRmMTNjLThmMzktNDI5MC1iNjlmLTdjNDhkZjViYjQwZSIsImRhdGEiOnt9LCJyYW5kb20iOiIxODUyZGI3OGQ3MTExYTMzMzc5OWI3MTQ1NTIwNmM3NyJ9.ugxg11ydCgGqNbeyGsHE7K8IOXRx0G8oLMKbcKUudWd6AJieTaXEi7J_AO7FsxVvHM9XC4BIvif4pMKBelxAeg

## Acknowledgements

S.V. acknowledges the Spanish Ministerio de Ciencia, Innovación y Universidades (MICIU) for projects PID2022-138265NA-I00 and CEX2021-001202-M. J.R-A. acknowledges the MICIU for projects PID2023-149691NB-I00 and CEX2021-001202-M, and the Generalitat de Catalunya (Project 2021SGR00354). The CSUC, the IQCTUB, the University of Barcelona, and the Red Española de Supercomputación (RES, project QH-2023-2-0004) are also acknowledged for computational resources. S.P.V. and K.S acknowledge support from 2 PSC-CUNY Awards, jointly funded by The Professional Staff Congress and The City University of New York (TRADA-48-433 and TRADB-51-383). S.P.V. received financial support from The Grace Spruch '47 Physics Fund at Brooklyn College. A portion of this research used resources at the Spallation Neutron Source, a DOE Office of Science User Facility operated by the Oak Ridge National Laboratory.

# -- Supporting Information --

## *Atomistic Description of Spin Crossover Under Pressure and its Giant Barocaloric Effect*

Sergi Vela,[a,b,*] Jordi Ribas-Arino,[a] Steven P. Vallone,[c,d] António M. dos Santos,[e] Malcolm A. Halcrow[f] and Karl Sandeman[c,d]

[a]Departament de Ciència de Materials i Química Física and IQTCUB, Universitat de Barcelona, Martí i Franquès 1, E-08028, Barcelona, Spain.
[b]Institut de Química Avançada de Catalunya (IQAC-CSIC), Barcelona, Spain. sergi.vela@iqac.csic.es

[c]Department of Physics, Brooklyn College of The City University of New York, 2900 Bedford Avenue Brooklyn NY 11210, USA
[d]Physics Program, The Graduate Center, CUNY, 365 Fifth Avenue NY 10016, USA

[e] Neutron Sciences Directorate, Oak Ridge National Laboratory, Oak Ridge, TN 37831-6393, USA

[f] School of Chemistry, University of Leeds, Woodhouse Lane, Leeds LS2 9JT, UK

**S1. Evolution of cell parameters and volume**

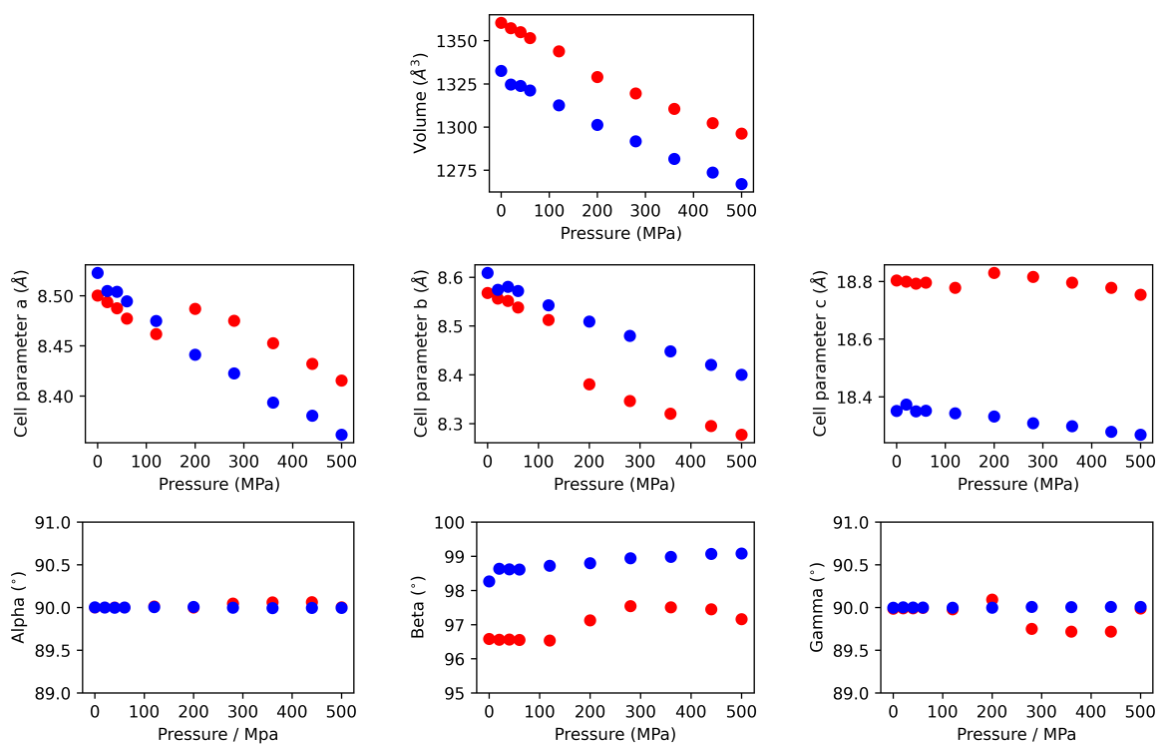

**Figure S1.** Evolution of volume and unit cell parameters with pressure, for the HT (red) and LT (blue) phases of **1**.

## S2. Evolution of intra- and inter-molecular structural descriptors

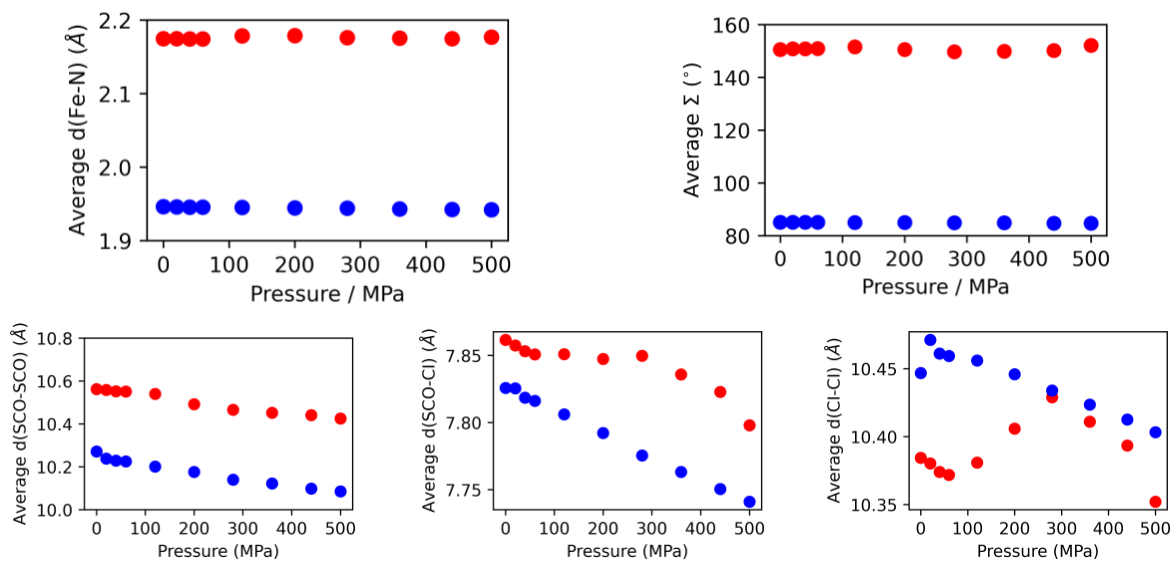

**Figure S2.** Evolution of (top) intramolecular Fe-N distance and Σ distortion parameter, and (bottom) intermolecular centroid-centroid distance between and among SCO and CI molecules with pressure, for the HT (red) and LT (blue) phases of **1**. Σ is defined as: $\Sigma = \sum_{i=1}^{12} |90 - \alpha_i|$, where $\alpha_i$ are the twelve *cis*-N–Fe–N angles about the iron atom.

## S3. Evolution of phonon frequencies

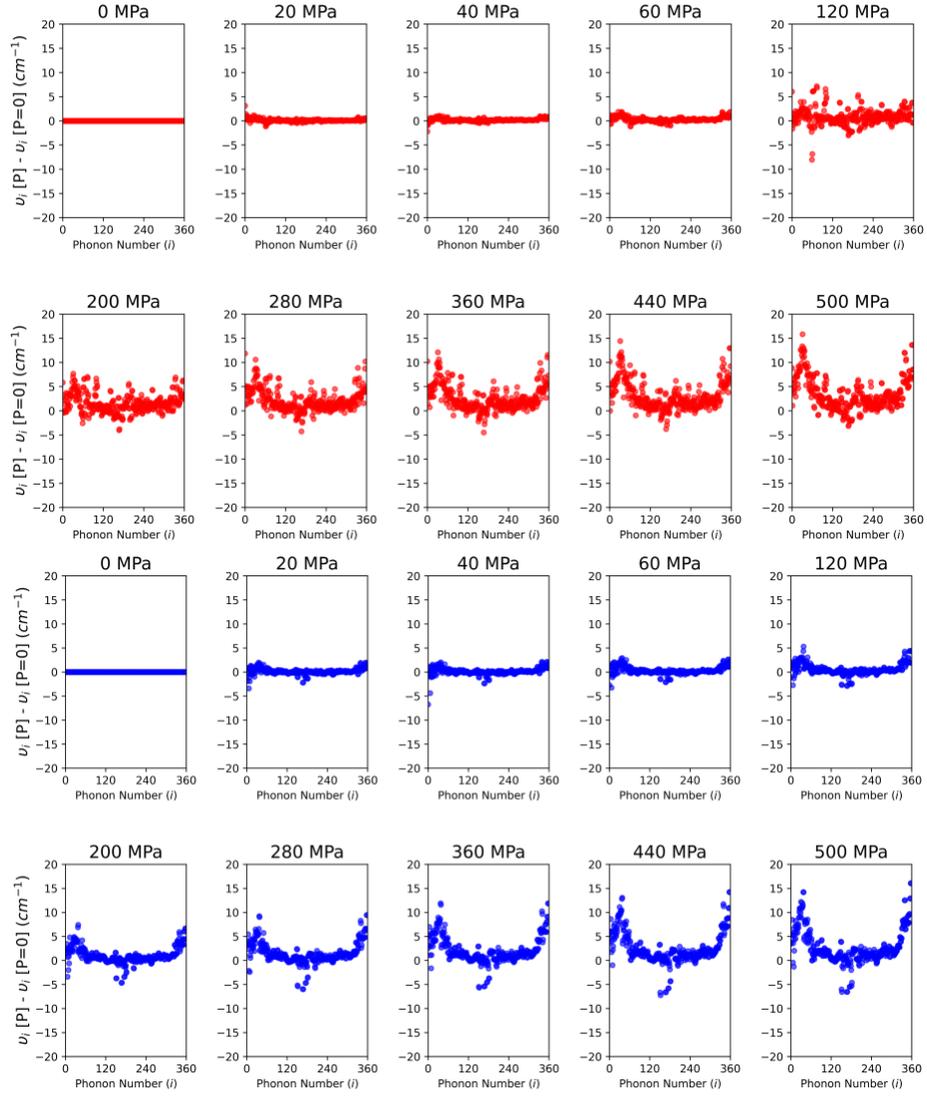

**Figure S3.** Evolution of HS (top, red) and LS (bottom, blue) phonon frequencies when increasing pressure. Each dot represents the difference between the frequency of a phonon at a given pressure $v_i[P]$ and at ambient pressure $v_i[P=0]$.

## S4. Evolution of phonon eigenvectors

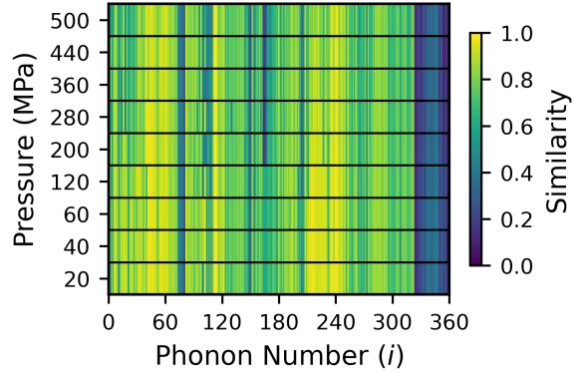

**Figure S4.** Similarity of LS phonons eigenvectors with respect to ambient pressure. See Figure 3 for HS and below for more details.

The similarity between two *i*-th eigenvectors has been computed as the normalized summation (over the *N* atoms) of cosine similarities between *j*-th atomic contributions to the eigenvector (vectors **A** and **B**). By taking the absolute value of the cosine similarity we ensure that an arbitrary sign change in the eigenvectors does not affect the metric (see Eq. S1). The unit cell of **1** contains 122 atoms, corresponding to 360 phonons (3N-6), each associated with an eigenvector with 122 atomic contributions.

$$\boldsymbol{sim_i} = \frac{\sum_j^N \left| \frac{A_j \cdot B_j}{\|A_j\| \|B_j\|} \right|}{N} = \frac{\sum_j^N |\cos(A_j B_j)|}{N} \qquad \text{Eq. S1}$$